\documentclass[10pt]{article}
\usepackage{titlesec}
\usepackage{amsmath,esdiff}
\usepackage{amssymb}
\usepackage{authblk}
\DeclareMathOperator{\csch}{cosech}
\DeclareMathOperator{\trace}{Tr}
\usepackage[normalem]{ulem}
\usepackage{bbold}
\usepackage[colorlinks=true,linkcolor=blue,citecolor=teal,urlcolor=teal]{hyperref}%
\usepackage{geometry}
\usepackage{setspace}
\usepackage{cite}
\usepackage{flushend}
\usepackage{mathrsfs}
\usepackage{hyperref}
\usepackage{graphicx}
\usepackage{cancel}
\usepackage{array,esint}
\usepackage[most]{tcolorbox}
\graphicspath{}
\titleformat{\section}{\fontsize{12}{12}\bfseries}{\thesection}{1em}{}
\twocolumn
\begin{document}
\twocolumn[\begin{@twocolumnfalse}
\title{\textbf{Equivalence principle and HBAR entropy of an atom falling into a quantum corrected black hole }}
\author{\textbf{Soham Sen${}^{a\dagger}$, Rituparna Mandal${}^{a*}$ and Sunandan Gangopadhyay${}^{a\ddagger}$}}
\affil{{${}^a$ Department of Theoretical Sciences}\\
{S.N. Bose National Centre for Basic Sciences}\\
{JD Block, Sector III, Salt Lake, Kolkata 700 106, India}}
\date{}
\maketitle
\begin{abstract}
\noindent In this work we have investigated the phenomenon of acceleration radiation exhibited by an atom falling into a quantum corrected Schwarzschild black hole.  We observe that the excitation-probability of the atom with simultaneous emission of a photon satisfies the equivalence principle when we compare it to the excitation probability of a mirror accelerating with respect to an atom. We also demonstrate the validity of the equivalence principle for a generic black hole geometry. Then we calculate the horizon brightened acceleration radiation (HBAR) entropy for this quantum corrected black hole geometry. We observed that the HBAR entropy has the form identical to that of Bekenstein-Hawking black hole entropy along with universal quantum gravity corrections.
\end{abstract}
\end{@twocolumnfalse}]
\section*{Introduction}
\let\thefootnote\relax\footnote{{}\\
{$\dagger$sensohomhary@gmail.com, soham.sen@bose.res.in}\\
{$*$drimit.ritu@gmail.com}\\
{$\ddagger$sunandan.gangopadhyay@gmail.com}}
\noindent The fact that radiation is emitted when atoms fall into a black hole has been an important area of research in recent times\cite{Fulling2}. The foundations of this problem lie in the general theory of relativity known to be the brain child of Albert Einstein \cite{Einstein15, Einstein16}. The remarkable observation made in the problem of atoms falling into a black hole is that the acceleration radiation emitted has a similar form to Hawking radiation to a distant observer, although it has a different form. The entropy of this radiation has been termed as horizon brightened acceleration radiation (HBAR) entropy\cite{Fulling2}. Indeed, such a result throws light on the connection between black hole physics and quantum optics\cite{Fulling2, Weiss, Philbin}. The study also gave an insight into the principle of  equivalence in general relativity. Historically, this connection was preceded by the link between gravitation, geometry and thermodynamics. The formulation of black hole thermodynamics \cite{Bekenstein, Bekenstein2, Hawking, Hawking2}, discovery of Hawking radiation \cite{Hawking, Hawking2}, particle emission from black holes \cite{Page, Page2, Page3}, the Unruh effect \cite{Unruh}, and 
%\textcolor{violet}
acceleration radiation \cite{Fulling21, Davies, DeWitt, Unruh2, Muller, Vanzella, Higuichi, Ordonez1, Ordonez2, Ordonez3, Ordonez4} reinforces this deep connection. 

\noindent Quantum mechanics on the other hand has been another pillar of theoretical physics. However, its unification with general relativity has proved to be a notorious problem. One of the main directions of research in theoretical physics has been to search for a unified theory of quantum gravity \cite{ROVELLI,CARLIP,ACV,KOPAPR}. An alternative approach to quantum gravity based on the asymptotic safety idea has also attached a lot of interest in the recent past.  This formalism is based on a scale dependent effective average action which describes all gravitational phenomena taking into account the effect of all loops\cite{Reuter0, Wetterich, ReuterWetterich, SReuter}. As a function of this scale, the effective average action satisfies a renormalization group equation. This then results in the flow of the Newton's gravitational constant as a function of the scale. An effective improved black hole solution was then obtained from the flow of the Newton's gravitational constant, whose metric reads \cite{Reuter}  
\begin{equation}\label{2.1}
ds^2=-f(r)dt^2+f(r)^{-1}dr^2+r^2d\theta^2+r^2\sin^2\theta d\phi^2
\end{equation}
where  
\begin{equation}\label{2.2}
f(r)=1-\frac{2G(r)M}{r}
\end{equation}
and $G(r)$ in natural units reads
\begin{equation}\label{2.2a}
G(r)=\frac{G }{1+\frac{\tilde{\omega} G}{r^2}}
\end{equation}
where $\tilde{\omega}$ is a constant that involves the quantum gravity correction to the black hole geometry coming from the renormalization group approach. Our plan in this work is to consider atoms falling into this black hole which incorporates quantum gravity corrections and draw conclusions from such a thought experiment. The aim is to look at two important physical issues. First of all, we would like to get an equivalent insight on the Einstein's principle of equivalence. There were several attempts to provide an alternative explanation of the equivalence principle \cite{Rohrlich}-\cite{SG}. Such an insight was obtained earlier in \cite{Fulling, SG} where the spontaneous excitation of a two-level atom in presence of a perfectly reflecting mirror in the generalized uncertainty principle framework was studied. It was shown in \cite{SG} that when the mirror is accelerating, the spatial oscillation of the probability of excitation of the atom gets modulated due to the generalized uncertainty principle, thereby leading to an explicit violation of the equivalence principle. Now atom falling into a black hole emits radiation which has exactly the same thermal spectrum, as that radiated by a fixed atom near an accelerating mirror. This can be then thought of as a different manifestation of the equivalence principle than the standard elevator description. Our goal is to look at the status of this principle when an atom falls into a renormalization group ``improved" black hole spacetime. The second aim is to look for quantum gravity corrections in the HBAR entropy and see whether they are logarithmic in nature similar to the corrections in the Bekenstein-Hawking entropy \cite{Bekenstein, Bekenstein2, Hawking, Hawking2, ParthaKaul}.
\section*{The quantum corrected black hole metric in Rindler form}
To begin our analysis, we start from the  Minkowski metric in 1+1-dimensional form
\begin{equation}\label{1.1}
ds^2=c^2dt^2-dz^2~.
\end{equation}
For a particle moving with a constant proper acceleration `$a$' in this flat spacetime, the time and position coordinates in terms of the proper time $\tau$ can be written as
\begin{align}
t(\tau)&=\frac{c}{a}\sinh\left(\frac{a\tau}{c}\right)\label{1.2}
\end{align}
\begin{align}
z(\tau)&=\frac{c^2}{a}\cosh\left(\frac{a\tau}{c}\right)~.\label{1.3}     
\end{align}
We shall now use the following coordinate transformations in eq.(\ref{1.1})
\begin{align}
t&=\frac{\rho}{c}\sinh(\frac{\tilde{a}\tilde{t}}{c})\label{1.4}\\
z&=\rho\cosh(\frac{\tilde{a}\tilde{t}}{c})~.\label{1.5}
\end{align}
Using eq.(s)(\ref{1.4},\ref{1.5}) in eq.(\ref{1.1}), the form of the metric gets modified as
\begin{equation}\label{1.6}
ds^2=\left(\frac{\tilde{a}\rho}{c^2}\right)^2c^2d\tilde{t}^2-d\rho^2~.
\end{equation}
Eq.(\ref{1.6}) is said to be the Rindler form of the metric describing constant acceleration of a particle. If we now compare eq.(\ref{1.4}) with eq.(\ref{1.2}), we see that the proper time of the particle takes the form
\begin{equation}\label{1.7}
\tau=\frac{\tilde{a}\tilde{t}}{a}~.
\end{equation} 
Comparing eq.(\ref{1.5}) with eq.(\ref{1.3}), we get the form of the uniform acceleration of the particle in Minkowski spacetime as
\begin{equation}\label{1.8}
a=\frac{c^2}{\rho}~.
\end{equation}
With this background, we shall now reduce the metric of the quantum corrected Schwarzschild black hole in the Rindler form to obtain the uniform acceleration of a freely falling particle close to the horizon. The horizons for the quantum corrected Schwarzschild black hole can be determined by setting $f(r)=0$ in eq.(\ref{2.2}). The forms of the outer and inner horizons read
\begin{equation}\label{2.3}
r_{\pm}=GM\pm \sqrt{G^2M^2-\tilde{\omega} G}~.
\end{equation}
In eq.(\ref{2.3}), $r_+$ and $r_-$ denotes the outer and inner horizons of the black hole. The outer horizon and $f(r)$ upto order $\tilde{\omega}$ (considering $\tilde{\omega}$ to be small) in natural units are given by 
\begin{align}
r_+&\approx 2GM-\frac{\tilde{\omega}}{2M}\label{2.4}~\\
f(r)&\approx 1-\frac{2 G M}{r}+\frac{2\tilde{\omega}G^2M}{r^3}\label{2.5}~.
\end{align}
We shall now carry out a near horizon expansion with respect to the outer horizon to obtain the Rindler form. The near horizon expansion can be obtained by making a Taylor series expansion of $f(r)$ about the outer horizon $r_+$. Keeping terms upto first order in the near horizon expansion parameter $(r-r_+)$, we get
\begin{equation}\label{2.6}
f(r)\cong f(r_+)+(r-r_+)\frac{df(r)}{dr}\biggr|_{r=r_+}=(r-r_+)f'(r_+)
\end{equation}
where we have used $f(r_+)=0$. 
Using this expansion, we obtain the line element in $1+1$-dimensions to be
\begin{equation}\label{2.6a}
\begin{split}
ds^2&=f(r)c^2dt^2-f(r)^{-1}dr^2\\
&\cong(r-r_+)f'(r_+)c^2dt^2-\frac{1}{(r-r_+)f'(r_+)}dr^2~.
\end{split}
\end{equation}
We now define a transformation of coordinates to obtain the Rindler form as follows
\begin{equation}\label{2.7}
\rho=2\sqrt{\frac{r-r_+}{f'(r_+)}}~.
\end{equation}
Using the above transformation in eq.(\ref{2.6a}), we obtain the Rindler form of the metric in $1+1$-dimensions as follows
\begin{equation}\label{2.8}
ds^2\cong \frac{\rho^2f'^2(r_+)}{4}c^2dt^2-d\rho^2~.
\end{equation}
Here $f'(r_+)$ upto order $\mathcal{O}(\tilde{\omega})$ reads
\begin{equation}\label{2.8a}
f'(r_+)=\frac{c^2}{2GM}-\frac{\hbar\tilde{\omega}c^3}{8G^2M^3}~.
\end{equation}
Comparing eq.(\ref{1.6}) with eq.(\ref{2.8}), we get the uniform acceleration corresponding to curves of constant $\rho$ upto linear order in $\tilde{\omega}$ as 
\begin{equation}\label{2.9}
\begin{split}
a&=\frac{c^2}{\rho}=\frac{c^2}{2\sqrt{\frac{r-r_+}{f'(r_+)}}}\\
&\cong\frac{c^2\sqrt{f'(r_+)}}{2\sqrt{r}}\left(1+\frac{r_+}{2r}\right)\\
&\cong\frac{c^3}{\sqrt{2GMr}}\left[1+\frac{GM}{rc^2}-\frac{\hbar\tilde{\omega c}}{8GM^2}\left(1+\frac{3GM}{c^2r}\right)\right]~.
\end{split}
\end{equation}
We shall use this result in the subsequent section to check the validity of the Einstein's equivalence principle.

\section*{Atom falling into the quantum corrected black hole}
In this section, we shall calculate the 	probability of transition from the ground state of a two-level atom falling into a quantum corrected Schwarzschild black hole to the excited state along with the emission of a photon.  We will denote the ground state and the excited state by $g$ and $e$. Before proceeding further, we note that in the quantum corrected black hole geometry the atom trajectory is given by the following equations
\begin{align}
\tau(r)&=-\int \frac{dr}{\sqrt{1-f(r)}}\nonumber\\
&\approx-\frac{1}{\sqrt{2GM}}\left(\frac{2}{3}r^{\frac{3}{2}}-\tilde{\omega}Gr^{-\frac{1}{2}}\right)+C'~\label{2.10}
\end{align}
\begin{align}
t(r)&=-\int\frac{dr}{f(r)\sqrt{1-f(r)}}\nonumber\\
\end{align}
\begin{align}
t(r)&\approx-4GM\left(\frac{r^{\frac{1}{2}}}{\sqrt{2GM}}+\frac{r^{\frac{3}{2}}}{3(2GM)^{\frac{3}{2}}}+\frac{1}{2}\ln\frac{\frac{\sqrt{r}}{\sqrt{2GM}}-1}{\frac{\sqrt{r}}{\sqrt{2GM}}+1}\right)\nonumber\\
&+\tilde{\omega}G\left(\frac{r^{\frac{1}{2}}/2GM}{(r-2GM)}+\frac{1}{2(2GM)^{\frac{3}{2}}}\ln\frac{\frac{\sqrt{r}}{\sqrt{2GM}}-1}{\frac{\sqrt{r}}{\sqrt{2GM}}+1}\right)+C\label{2.11}
\end{align}
where $C'$ and $C$ are integration constants. For simplicity we shall set $2GM=1$ during the calculation of the excitation probability of the atom with simultaneous photon emission.  For a massless scalar photon with wave function $\Psi$, the covariant Klein Gordon equation is given as
\begin{equation}\label{2.11a}
\frac{1}{\sqrt{-g}}\partial_\mu\left(\sqrt{-g}g^{\mu\nu}\partial_\nu\right)\Psi=0~.
\end{equation}
 In the $s-$wave approximation eq.(\ref{2.11a}) takes the form
\begin{equation}\label{2.12}
\frac{1}{T(t)}\frac{d^2T(t)}{dt^2}-\frac{f(r)}{r^2R(r)}\frac{d}{dr}\left(r^2f(r)\frac{dR(r)}{dr}\right)=0
\end{equation}
where the wave function $\Psi(t,r)$ of the scalar photon is written as
\begin{equation}\label{2.13}
\Psi(t,r)=T(t)R(r)~.
\end{equation}
The general solution of eq.(\ref{2.12}) reads
\begin{equation}\label{2.14}
\Psi_\nu(t,r)=\exp\left[- i\nu t+ i\nu\int\frac{dr}{f(r)}\right]~.
\end{equation}
In eq.(\ref{2.14}), `$\nu$' describes the photon frequency which is observed by an observer sitting at infinity from the quantum corrected black hole.  In this work, we follow the model in \cite{Fulling2} consisting of a mirror surrounding the black hole to shield Hawking radiation coming out of it. This set up is identical to that of a Boulware vacuum\cite{Boulware}. Now defining an operator $\hat{\zeta}=|g\rangle\langle e|$, the atom field interaction Hamiltonian can be written as 
\begin{equation}\label{2.15}
\hat{H}_I(\tau)=\hbar \mathcal{G}[\hat{b}_\nu \Psi_\nu(t(\tau),r(\tau))+H.c.][\hat{\zeta}e^{-i\Omega\tau}+h.c.]~.
\end{equation}
In eq.(\ref{2.15}), $\mathcal{G}$ is the atom-field coupling constant, $\hat{b}_\nu$ is the annihilation operator of the photon and $\Omega$ is the atomic frequency. With the simultaneous emission of a scalar photon, the probability of excitation of the atom is given by
\begin{equation}\label{2.17}
\begin{split}
P_{g,0\rightarrow e,1}&=\frac{1}{\hbar^2}\left|\int d\tau\langle 1_\nu,e|\hat{H}_{I}(\tau)|0_\nu,g\rangle\right|^2\\
&=\mathcal{G}^2\left|\int d\tau e^{i\nu t(\tau)-i\nu r_*(\tau)}e^{i\Omega\tau}\right|^2\\ 
&=\mathcal{G}^2\left|\int_{\infty}^{1-\frac{\tilde{\omega}}{2M}}dr \left(\frac{d\tau}{dr}\right) e^{i\nu t(r)-i\nu r_*(r)}e^{i\Omega\tau(r)}\right|^2\\
&\approx\mathcal{G}^2\left|\int_1^\infty dr\sqrt{r}\left(1+\frac{\tilde{\omega}}{4Mr^2}\right)\Psi^*_\nu(r)e^{i\Omega\tau(r)}\right|^2~
\end{split}
\end{equation}
where the forms of $\tau(r)$ and $t(r)$ are given in eq.(s)(\ref{2.10},\ref{2.11}). We will now make a change of variables as follows 
\begin{equation}\label{2.18}
y=\frac{2\Omega}{3}(r^{\frac{3}{2}}-1)~
\end{equation}
where $\Omega\gg 1$.
Using eq.(\ref{2.18}) in eq.(\ref{2.17}), the form of $P_{g,0\rightarrow e,1}$ is obtained as 
\begin{equation}\label{2.19}
P_{g,0\rightarrow e,1}=\frac{\mathcal{G}^2}{\Omega^2}\biggr|\int_0^\infty dy\mathcal{A}_0e^{-i\nu \mathcal{A}_1(y)}e^{-i\Omega \mathcal{A}_2(y)}\biggr|^2
\end{equation}
where
\begin{align}
\mathcal{A}_0&=\left(1+\frac{\tilde{\omega}}{4M}\left(1+\frac{3y}{2\Omega}\right)^{-\frac{4}{3}}\right)\label{2.20}
\end{align}
\begin{align}
\mathcal{A}_1(y)&=2\left(1+\frac{3y}{2\Omega}\right)^{\frac{1}{3}}+\frac{2}{3}\left(1+\frac{3y}{2\Omega}\right)+\left(1+\frac{3y}{2\Omega}\right)^{\frac{2}{3}}\nonumber\\
&+2\ln\biggr[\left(1+\frac{3y}{2\Omega}\right)^{\frac{1}{3}}-1\biggr]+\frac{\tilde{\omega}}{2M}\biggr\{-\ln\left[1+\frac{3y}{2\Omega}\right]^{\frac{2}{3}}\nonumber\\
&+2\ln\biggr[\left(1+\frac{3y}{2\Omega}\right)^{\frac{1}{3}}-1\biggr]+\frac{1}{\left(1+\frac{3y}{2\Omega}\right)^{\frac{1}{3}}-1}\biggr\}\label{2.21}\\
\mathcal{A}_2(y)&=\frac{2}{3}\left(1+\frac{3y}{2\Omega}\right)-\frac{\tilde{\omega}}{2M}\left(1+\frac{3y}{2\Omega}\right)^{-\frac{1}{3}}\label{2.22}~.
\end{align}
Simplify eq.(\ref{2.19}), $P_{g,0\rightarrow e,1}$ upto linear order in $\tilde{\omega}$ reads
\begin{align}
P_{g,0\rightarrow e,1}\approx&\frac{\mathcal{G}^2}{\Omega^2}\biggr|\int_0^\infty dy\left(1+\frac{\tilde{\omega}}{4M}\right)y^{-\left(2i\nu+\frac{i\tilde{\omega}\nu}{M}\right)}\nonumber\\
&e^{-iz\left(1+\frac{3\nu}{\Omega}+\frac{\tilde{\omega}}{4M}\left(1-\frac{2\nu}{\Omega}\right)\right)}\left(1-\frac{i\tilde{\omega}\nu\Omega}{yM}\right)\biggr|^2~.\label{2.23}
\end{align}
We now make another change of coordinates as follows
\begin{equation}\label{2.24}
z=y\left(1+\frac{3\nu}{\Omega}+\frac{\tilde{\omega}}{4M}\left(1-\frac{2\nu}{\Omega}\right)\right)~.
\end{equation}
Using the above coordinate transformation in eq.(\ref{2.23}), we get
\begin{align}
P_{g,0\rightarrow e,1}\approx&\frac{\mathcal{G}^2\left(1+\frac{\tilde{\omega}}{2M}\frac{5\nu}{\Omega+3\nu}\right)}{\Omega^2\left(1+\frac{3\nu}{\Omega}\right)^2}\biggr|\int_0^\infty dz ~e^{-iz}z^{-\left(2i\nu+\frac{i\tilde{\omega}\nu}{M}\right)}\nonumber\\
&\left(1-\frac{i\tilde{\omega}\nu\Omega}{z m}\left(1+\frac{3\nu}{\Omega}\right)\right)\biggr|^2\nonumber\\
\approx& \frac{4\pi \mathcal{G}^2\nu\left(1+\frac{\tilde{\omega}}{2M}\frac{\Omega+8\nu}{\Omega+3\nu}\right)}{\Omega^2\left(1+\frac{3\nu}{\Omega}\right)^2}\frac{1}{e^{4\pi\nu\left(1+\frac{\tilde{\omega}}{2M}\right)}-1}~.\label{2.26}
\end{align} 
The absorption probability on the other hand is given by
\begin{equation}\label{2.27}
P_{e,0\rightarrow g,1}\approx \frac{4\pi \mathcal{G}^2\nu\left(1+\frac{\tilde{\omega}}{2M}\frac{\Omega-8\nu}{\Omega-3\nu}\right)}{\Omega^2\left(1-\frac{3\nu}{\Omega}\right)^2}\frac{1}{1-e^{-4\pi\nu\left(1+\frac{\tilde{\omega}}{2M}\right)}}~.
\end{equation}
It can be observed that for a very high photon frequency, the excitation probability becomes considerably smaller. The atomic frequency is considerably higher and in our case we consider $\Omega\gg\nu$. With dimensional reconstruction and considering the case when atomic frequency is substantially higher than the emitted photon frequency, we obtain the excitation probability to be
\begin{align}
P_{g,0\rightarrow e,1}\approx& \frac{4\pi \mathcal{G}^2\nu\left(\frac{2GM}{c^3}+\frac{\hbar\tilde{\omega}}{2Mc^2}\right)}{\Omega^2}\frac{1}{e^{4\pi\nu\left(\frac{2GM}{c^3}+\frac{\hbar\tilde{\omega}}{2Mc^2}\right)}-1}~.\label{2.28}
\end{align} 
Now for the mirror-atom system\cite{Fulling,Fulling2}, the excitation probability for the case when the mirror is accelerating with respect to the fixed atom, reads
\begin{equation}\label{2.29}
P^{AM}_{g,0\rightarrow e,1}=\frac{4\pi\mathcal{G}^2\nu c}{a\Omega^2}\frac{1}{e^{\frac{2\pi c\nu}{a}}-1}~.
\end{equation}
It is observed in \cite{Fulling2} that the Einstein principle of equivalence holds for the case of a Schwarzschild black hole. In this work, we shall check its status for the quantum corrected Schwarzschild black hole. At first we replace $\nu$ by $\nu_\infty$ (photon frequency observed by a distant observer) in eq.(\ref{2.28}), where the relation between $\nu$ and $\nu_\infty$ is connected by the gravitational red-shift factor as
\begin{equation}\label{2.30}
\nu=\frac{\nu_\infty}{\sqrt{f(r)}}\implies \nu_\infty\cong \nu \sqrt{(r-r_+)f'(r_+)}
\end{equation}
where we have substituted $f(r)$ from eq.(\ref{2.6}) to obtain the analytical form of $\nu_\infty$. From the first line of eq.(\ref{2.9}) we can write
\begin{equation}\label{2.30a}
\begin{split}
\sqrt{r-r_+}&=\frac{c^2}{2a}\sqrt{f'(r_+)}\\
\implies \nu\sqrt{(r-r_+)f'(r_+)}&=\nu_\infty=\nu f'(r_+)\frac{c^2}{2a}\\
\implies \nu_\infty&=\frac{\nu c^2}{2a}\left[\frac{c^2}{2GM}-\frac{\hbar\tilde{\omega}c^3}{8G^2M^3}\right]
\end{split}
\end{equation}
\noindent where we have used eq.(\ref{2.8a}) to replace $f'(r_+)$ in the last line of the above equation with proper dimensional reconstruction. For a distant observer the excitation probability in eq.(\ref{2.28}) takes the following form 
\begin{equation}\label{2.30b}
\begin{split}
\mathcal{P}\biggr|_{\nu=\nu_\infty}&\cong \frac{4\pi \mathcal{G}^2\nu_\infty\left(\frac{2GM}{c^3}+\frac{\hbar\tilde{\omega}}{2Mc^2}\right)}{\Omega^2}\frac{1}{e^{4\pi\nu_\infty\left(\frac{2GM}{c^3}+\frac{\hbar\tilde{\omega}}{2Mc^2}\right)}-1}
\end{split}
\end{equation}
where
\begin{equation}\label{2.30c}
\begin{split}
\nu_\infty\left(\frac{2GM}{c^3}+\frac{\hbar\tilde{\omega}}{2Mc^2}\right)=&\frac{\nu c^2}{2a}\left[\frac{c^2}{2GM}-\frac{\hbar\tilde{\omega}c^3}{8G^2M^3}\right]\\ &\times\left[\frac{2GM}{c^3}+\frac{\hbar\tilde{\omega}}{2Mc^2}\right]\\
=&\frac{\nu c}{2 a}\left(1+\mathcal{O}(\tilde{\omega}^2)\right)\cong\frac{\nu c}{2a}.
\end{split}
\end{equation}
Using relation (\ref{2.30c}) in eq.(\ref{2.30b}), we obtain the probability as follows
\begin{equation}\label{2.30d}
\begin{split}
\mathcal{P}=\frac{2\pi\mathcal{G}^2\nu c}{a\Omega^2}\frac{1}{e^{\frac{2\pi c\nu}{a}}-1}~.
\end{split}
\end{equation}
The excitation probability in eq.(\ref{2.30d}) is identical to eq.(\ref{2.29}) upto a constant factor. This particular insight into the equivalence principle has a very subtle difference than the usual description of the equivalence principle using the elevator description. In the case of the accelerating mirror, the normal modes of the fields get modified and for the freely falling atom the gravitational field of the black hole is responsible for the same effect. What we observe interestingly is that the emitted radiation is identical for the two cases. This implies that the effect of the mirror acceleration and that of gravitational field is same on the atom. This result indicates that the aforementioned description of the equivalence principle holds exactly in case of this renormalization group improved Schwarzschild black hole as well. 

\noindent The absorption probability in eq.(\ref{2.27}) for $\Omega\gg\nu$ reads
\begin{equation}\label{2.31}
P_{e,0\rightarrow g,1}\approx \frac{4\pi \mathcal{G}^2\nu\left(\frac{2GM}{c^3}+\frac{\hbar\tilde{\omega}}{2Mc^2}\right)}{\Omega^2}\frac{1}{1-e^{-4\pi\nu\left(\frac{2GM}{c^3}+\frac{\hbar\tilde{\omega}}{2Mc^2}\right)}}~.
\end{equation}
In the next section, we shall use this result to compute the HBAR entropy.
\section*{Atom falling into a generic black hole spacetime}
The previous sections dealt with the case of a quantum corrected black hole. We will now consider a generic  black hole geometry with lapse function $f(r)$ having an event horizon at $r_+$. The excitation probability is given as follows
\begin{equation}\label{3.1}
\begin{split}
\mathcal{P}_{f,r_+}=&\mathcal{G}^2\left|\int_\infty^{r_+} dr \frac{d\tau}{dr}e^{i\nu t(r)-i\nu r_*(r)}e^{i\Omega\tau(r)}\right|^2\\
=&\mathcal{G}^2\biggr|\int_{r_+}^{\infty}dr\frac{1}{\sqrt{1-f(r)}}e^{-i\nu\int\frac{dr}{f(r)\sqrt{1-f(r)}}-i\nu\int\frac{dr}{f(r)}}\\
&\times e^{-i\Omega\int\frac{dr}{\sqrt{1-f(r)}}}\biggr|^2~.
\end{split}
\end{equation}
We have used the form of $\tau(r)$, $t(r)$ and $r_*(r)$ to obtain the form in the above line of eq.(\ref{3.1}).Using the near horizon expansion of $f(r)$ from eq.(\ref{2.6}) in the above equation, we obtain the following expression for the excitation probability
\begin{equation}\label{3.2}
\begin{split}
\mathcal{P}_{f,r_+}=&\mathcal{G}^2\biggr|\int_{r_+}^{\infty}dr\frac{e^{-i\nu\int\frac{dr}{(r-r_+)f'(r_+)\sqrt{1-(r-r_+)f'(r_+)}}}}{\sqrt{1-(r-r_+)f'(r_+)}}\\
&\times e^{-i\nu\int\frac{dr}{(r-r_+)f'(r_+)}} e^{-i\Omega\int\frac{dr}{\sqrt{1-(r-r_+)f'(r_+)}}}\biggr|^2\\
=&\mathcal{G}^2\biggr|\int_{r_+}^{\infty}dr\frac{e^{-\frac{2i\nu}{f'(r_+)}\ln\left(1-\sqrt{1-(r-r_+)f'(r_+)}\right)}}{\sqrt{1-(r-r_+)f'(r_+)}}\\
&\times e^{\frac{2i\Omega}{f'(r_+)}\sqrt{1-(r-r_+)f'(r_+)}}\biggr|^2~.
\end{split}
\end{equation}
We will now make a change of variables as follows
\begin{equation}\label{3.3}
r-r_+=\frac{\kappa}{\Omega}
\end{equation}
where $f'(r_+),\nu,\kappa\ll \Omega$ (in natural units).
Using the above change in variables from eq.(\ref{3.3}) in eq.(\ref{3.2}), we obtain the following form of the excitation probability
\begin{equation}\label{3.4}
\begin{split}
\mathcal{P}_{f,r_+}\cong &\frac{\mathcal{G}^2}{\Omega^2}\biggr|\int_0^\infty d\kappa \left(1+\frac{\kappa}{2\Omega}f'(r_+)\right)e^{-\frac{2i\nu}{f'(r_+)}\ln\left(\frac{\kappa}{2\Omega}f'(r_+)\right)}\\
&\times e^{\frac{2i\Omega}{f'(r_+)}\left(1-\frac{\kappa}{2\Omega}f'(r_+)\right)}\biggr|^2\\
=&\frac{\mathcal{G}^2}{\Omega^2}\biggr|\int_0^\infty d\kappa \left(1+\frac{\kappa}{2\Omega}f'(r_+)\right)\kappa^{-\frac{2i\nu}{f'(r_+)}}e^{-i\kappa}\biggr|^2\\
=&\frac{4\pi\mathcal{G}^2\nu}{f'(r_+)\Omega^2}\left(\left(1-\frac{\nu}{\Omega}\right)^2+\frac{f'^2(r_+)}{4\Omega^2}\right)\frac{1}{e^{\frac{4\pi\nu}{f'(r_+)}}-1}~.
\end{split}
\end{equation}
As $\nu,f'(r_+)\ll \Omega$ (in natural units), the probability can be recast in the following form  
\begin{equation}\label{3.5}
\mathcal{P}_{f,r_+}\cong \frac{4\pi\mathcal{G}^2\nu}{f'(r_+)\Omega^2}\frac{1}{e^{\frac{4\pi\nu}{f'(r_+)}}-1}~.
\end{equation}
We can again recast the probability in terms of $\nu_\infty$ as follows (as was done earlier in eq.(\ref{2.30b}))
\begin{equation*}
\begin{split}
\mathcal{P}_{f,r_+}=&\frac{4\pi\mathcal{G}^2\nu_\infty}{f'(r_+)\Omega^2}\frac{1}{e^{\frac{4\pi\nu_\infty}{f'(r_+)}}-1}
\end{split}
\end{equation*}
\begin{equation}\label{3.6}
\begin{split}
=&\frac{4\pi\mathcal{G}^2\nu \sqrt{(r-r_+)f'(r_+)}}{f'(r_+)\Omega^2}\frac{1}{e^{\frac{4\pi\nu \sqrt{(r-r_+)f'(r_+)}}{f'(r_+)}}-1}~.
\end{split}
\end{equation}
Using eq.(\ref{2.30a}) in eq.(\ref{3.6}), we get the following form of the probability
\begin{equation}\label{3.7}
\begin{split}
\mathcal{P}_{f,r_+}=&\frac{4\pi\mathcal{G}^2\left(\frac{\nu}{2a}f'(r_+)\right)}{f'(r_+)\Omega^2}\frac{1}{e^{\frac{4\pi\left(\frac{\nu}{2a}f'(r_+)\right)}{f'(r_+)}}-1}\\
=&\frac{2\pi\mathcal{G}^2\nu}{a\Omega^2}\frac{1}{e^{\frac{2\pi\nu}{a}}-1}~.
\end{split}
\end{equation}
With proper dimensional reconstruction, eq.(\ref{3.7}) can be rewritten  as follows
\begin{equation}\label{3.8}
\mathcal{P}_{f,r_+}=\frac{2\pi\mathcal{G}^2\nu c}{a\Omega^2}\frac{1}{e^{\frac{2\pi\nu c}{a}}-1}~.
\end{equation}
Hence, following the same arguments after eq.(\ref{2.30d}), we can conclude that the equivalence principle holds for a generic black hole spacetime.

%%%%%%%%%%%%%%%%%%%%%%%%%%%%%%%%%%%%%%%%%%%%%%%
%%%%%%%%%%%%%%%%%%%%%%%%%%%%%%%%%%%%%%%%%%%%%%%

\section*{Modified HBAR entropy}
The term \textit{horizon brightened acceleration radiation entropy} (HBAR entropy) was first introduced in \cite{Fulling2}. In this section we will calculate the HBAR entropy for the quantum corrected black hole background. We are considering a case when two-level atoms with transition frequency $\Omega$ are falling into the event horizon of the black hole at a rate $\kappa$.  In this work we have used the quantum statistical approach to determine the entropy and hence we have used the density matrix formalism. If the microscopic change in the field density matrix is $\delta \rho_a$ for one atom then the overall macroscopic change in the field density matrix for $\Delta \mathcal{N}$ atoms is given by \cite{Fulling2}
\begin{equation}\label{2.32}
\Delta \rho=\sum\limits_a \delta\rho_a=\Delta \mathcal{N}\delta\rho
\end{equation} 
where
\begin{equation}\label{2.33}
\frac{\Delta \mathcal{N}}{\Delta t}=\kappa~.
\end{equation}
Putting $\Delta\mathcal{N}$ in eq.(\ref{2.33}), we obtain
\begin{equation}\label{2.34}
\frac{\Delta \rho}{\Delta t}=\kappa\delta\rho~.
\end{equation}
We now use the Lindblad master equation for the density matrix given by 
\begin{equation}\label{2.35}
\begin{split}
\frac{d\rho}{dt}=&-\frac{\Gamma_{abs}}{2}\left(\rho b^\dagger b+b^\dagger b \rho-2b\rho b^\dagger\right)\\
&-\frac{\Gamma_{exc}}{2}\left(\rho b b^\dagger+b b^\dagger \rho-2b^\dagger\rho b\right)
\end{split}
\end{equation}
where $\Gamma_{exc}$ is the excitation rate and $\Gamma_{abs}$ is the absorption rate given by $\Gamma_{exc/abs}=\kappa P_{exc/abs}$ with $P_{exc/abs}$ 
being given by eq.(s)(\ref{2.26},\ref{2.27}).
Taking the expectation of eq.(\ref{2.35}) with respect to some arbitrary state $|n\rangle$, we obtain
\begin{equation}\label{2.36}
\begin{split}
\dot{\rho}_{n,n}=&-\Gamma_{abs}\left(n\rho_{n,n}-(n+1)\rho_{n+1,n+1}\right)\\&-\Gamma_{exc}\left((n+1)\rho_{n,n}-n\rho_{n-1,n-1}\right)~.
\end{split}
\end{equation}
The steady state solution is now used to obtain the HBAR entropy. So we set $\dot{\rho}_{n,n}=0$ in eq.(\ref{2.36}) and for $n=0$ we obtain the relation between $\rho_{1,1}$ and $\rho_{0,0}$ as follows
\begin{equation}\label{2.39}
\rho_{1,1}=\frac{\Gamma_{exc}}{\Gamma_{abs}}\rho_{0,0}~.
\end{equation}  
Repeating this procedure, we finally obtain
\begin{equation}\label{2.40}
\rho_{n,n}=\left(\frac{\Gamma_{exc}}{\Gamma_{abs}}\right)^n\rho_{0,0}~.
\end{equation}
To obtain $\rho_{0,0}$ in the above relation, we use the condition $\trace(\rho)=1$. This gives
\begin{align}
\sum\limits_n \rho_{n,n}&=1\nonumber\implies \rho_{0,0}\sum_n\left(\frac{\Gamma_{exc}}{\Gamma_{abs}}\right)^n=1\nonumber\\
\implies \rho_{0,0}&=1-\frac{\Gamma_{exc}}{\Gamma_{abs}}\label{2.41}~.
\end{align}
Using $\rho_{0,0}$ from the above equation in eq.(\ref{2.40}), we obtain the steady state solution of the density matrix to be
\begin{equation}\label{2.42}
\rho_{n,n}^\mathcal{S}=\left(\frac{\Gamma_{exc}}{\Gamma_{abs}}\right)^n\left(1-\frac{\Gamma_{exc}}{\Gamma_{abs}}\right)
\end{equation}
where $\frac{\Gamma_{exc}}{\Gamma_{abs}}$ is given by (in the approximation $\Omega\gg\nu$)
\begin{equation}\label{2.43}
\frac{\Gamma_{exc}}{\Gamma_{abs}}\approx \left(1+\frac{\hbar\tilde{\omega}}{\mathcal{R}Mc}\frac{5\nu}{\Omega}\right)e^{-4\pi\nu\left(\frac{\mathcal{R}}{c}+\frac{\hbar\tilde{\omega}}{2Mc^2}\right)}~.
\end{equation}
Here, $\mathcal{R}=\frac{2GM}{c^2}$. The von-Neumann entropy for the system is given by
\begin{equation}\label{2.44}
S_\rho=-k_B\sum\limits_{n,\nu}\rho_{n,n}\ln(\rho_{n,n})
\end{equation}
and the rate of change of entropy due to the generation of real photons is obtained as
\begin{equation}\label{2.45}
\dot{S_\rho}=-k_B\sum\limits_{n,\nu}\dot{\rho}_{n,n}\ln(\rho_{n,n})~.
\end{equation}
The rate of change of the entropy, using the steady state density matrix solution, is given as
\begin{equation}\label{2.46}
\dot{S_\rho}\approx-k_B\sum\limits_{n,\nu}\dot{\rho}_{n,n}\ln(\rho^{\mathcal{S}}_{n,n})~.
\end{equation}
Using the form of the density matrix $\rho_{n,n}^{\mathcal{S}}$ in eq.(\ref{2.42}), we obtain
\begin{equation}\label{2.47}
\begin{split}
\dot{S}_\rho=&4\pi k_B\left(\frac{\mathcal{R}}{c}+\frac{\hbar\tilde{\omega}}{2Mc^2}\right)\sum\limits_\nu\left(\sum\limits_n n\dot{\rho}_{n,n}\right)\nu\\&-k_B\sum\limits_\nu\left(\sum\limits_nn\dot{\rho}_{n,n}\right)\ln\left[1+\frac{\hbar\tilde{\omega}}{\mathcal{R}Mc}\frac{5\nu}{\Omega}\right]\\
\approx&4\pi k_B\left(\frac{\mathcal{R}}{c}+\frac{\hbar\tilde{\omega}}{2Mc^2}\right)\sum\limits_\nu\dot{\bar{n}}_\nu\nu-\frac{5\hbar\tilde{\omega}k_B}{\mathcal{R}\Omega Mc}\sum\limits_\nu\dot{\bar{n}}_\nu\nu~.
\end{split}
\end{equation}
 In eq.(\ref{2.47}), $\dot{\bar{n}}_\nu$ is the flux due to photons from the atoms freely falling in the black hole and the total rate of energy loss due to emitted photons is $\hbar\sum\limits_\nu\dot{\bar{n}}_\nu\nu=\dot{m}_pc^2$. The black hole area in our case is given by
\begin{equation}\label{2.48}
\begin{split}
A_{Qbh}&=4\pi r_+^2\\
&\approx\frac{16\pi G^2M^2}{c^4}-\frac{8\pi\hbar\tilde{\omega}G}{c^3}\end{split}~.
\end{equation}  
Taking time derivative of the both sides in eq.(\ref{2.48}), we get
\begin{equation}\label{2.49}
\dot{A}_{Qbh}=\frac{32\pi G^2M\dot{M}}{c^4}
\end{equation}
where $\dot{M}=\dot{m}_p+\dot{m}_{atom}$. We now define the rate of change in the black hole area due to emitting photons as follows
\begin{equation}\label{2.50}
\dot{A}_p=\frac{32\pi G^2M\dot{m}_p}{c^4}~.
\end{equation}
From these results we can infer that when no atoms are falling in the black hole, then $A_p$ should be equivalent to the area of the black hole and $A_{atom}$ should be zero. A better way to understand the above result is as follows. Before the atom crosses the black hole event horizon, that is, before the atom contributes to the black hole mass, the freely falling atom emits HBAR radiation.  Therefore, one can separate in time the change of the black hole entropy associated with HBAR radiation from an atom and that associated with that atom's mass.

\noindent The final form of $\dot{S}_\rho$ in terms of $A_p$ can therefore be written as
\begin{equation}\label{2.51}
\begin{split}
\dot{S}_\rho\approx&\frac{k_Bc^3}{4\hbar G}\dot{A}_p+\tilde{\omega}\pi k_B\frac{d}{d t}(\ln A_p)+\frac{10\tilde{\omega}\sqrt{\pi}k_Bc}{\Omega}\frac{d}{d t}\left(A_p^{-\frac{1}{2}}\right)\\
=&\frac{d}{d t}\left(\frac{k_Bc^3}{4\hbar G}A_p+\tilde{\omega}\pi k_B\ln A_p+\frac{10\tilde{\omega}\sqrt{\pi}k_Bc}{\Omega}A_p^{-\frac{1}{2}}\right)~.
\end{split}
\end{equation}
We find that the horizon brightened acceleration radiation entropy for the case of a simple Schwarzschild black hole \cite{Fulling2} gets modified with the time derivatives of a log term and an inverse square root term due to quantum gravity corrections introduced in the Schwarzschild black hole. This indicates that logarithmic corrections also appear in HBAR entropy just as it appears in usual Bekenstein-Hawking black hole entropy \cite{Bekenstein, Bekenstein2, Hawking, Hawking2, ParthaKaul}. 
\section*{Conclusion}
In this work we have considered the case of a two-level atom freely falling near the outer horizon of a  renormalization group improved Schwarzschild black hole. The entire system is analysed by considering that the outer horizon of the black hole is surrounded with a mirror to prevent infalling atoms from interacting with the Hawking radiation. Further, the mirror ensures that the initial state of the field appears vacuum-like in the reference frame of the external observer. Our investigation in this context involved the computation of the atom-field excitation probability and the calculation of the rate of change of the horizon brightened acceleration radiation entropy or the HBAR entropy. The excitation probability obtained in this work involves a Planck like factor indicating emission of real photons in the virtual process. Interestingly, we find that even with quantum gravity corrections, the Einstein equivalence principle holds indicating that the principle may be a fundamental reality even in a quantum gravity setting. We then extended this calculation for a generic black hole metric and observed that the equivalence principle holds in a general setting as well. For the next part of the work we have calculated the HBAR entropy using the quantum optics approach\cite{Fulling2} and we have observed that the HBAR entropy has a Bekenstein-Hawking like entropy term along with a logarithmic and inverse square root quantum gravity term. The nature of the corrections appearing in the HBAR entropy are remarkably the same as those obtained in case of usual Bekenstein-Hawking entropy of a black hole.


\begin{thebibliography}{8}
\bibitem{Fulling2}
M. O. Scully, S. A. Fulling, D. M. Lee, D. N. Page, W. P. Schleich and A. A. Svidzinsky, \href{https://doi.org/10.1073/pnas.1807703115}{Proc. Natl. Acad. Sci. U.S.A. 115 (2018) 8131}.
\bibitem{Einstein15}
A. Einstein, Sitzungsber Preuss Akad Wiss (1915) 844.
\bibitem{Einstein16}
A. Einstein, \href{https://doi.org/10.1002/andp.19163540702}{Ann. der Physik 49 (1916) 769}.
\bibitem{Weiss}
P. Weiss, \href{https://www.sciencenews.org/article/black-hole-recipe-slow-light-swirl-atoms}{Sci. News 157 (2000) 86}.
\bibitem{Philbin}
T. G. Philbin, et al., \href{http://dx.doi.org/10.1126/science.1153625}{Science 319 (2008) 1367}.
\bibitem{Bekenstein}
J.D. Bekenstein, \href{https://doi.org/10.1007/BF02757029}{Lett. Nuovo Cimento 4 (1972) 737}.
\bibitem{Bekenstein2}
J. D. Bekenstein, \href{https://link.aps.org/doi/10.1103/PhysRevD.7.2333}{Phys. Rev. D 7 (1973) 2333}.
\bibitem{Hawking}
S.W. Hawking, \href{https://doi.org/10.1038/248030a0}{Nature 248 (1974) 30}.
\bibitem{Hawking2}
S.W. Hawking, \href{https://doi.org/10.1007/BF02345020}{Commun. Math. Phys 43 (1975) 199}.
\bibitem{Page}
D. N. Page, \href{https://link.aps.org/doi/10.1103/PhysRevD.13.198}{Phys. Rev. D 13 (1976) 198}.
\bibitem{Page2}
D. N. Page, \href{https://link.aps.org/doi/10.1103/PhysRevD.14.3260}{Phys. Rev. D 14 (1976) 3260}.
\bibitem{Page3}
D. N. Page, \href{https://link.aps.org/doi/10.1103/PhysRevD.16.2402}{Phys. Rev. D 16 (1977) 2402}.
\bibitem{Unruh}
W. G. Unruh, \href{https://link.aps.org/doi/10.1103/PhysRevD.14.870}{Phys. Rev. D 14 (1976) 870}.
\bibitem{Fulling21}
S. A. Fulling, \href{https://link.aps.org/doi/10.1103/PhysRevD.7.2850}{Phys. Rev. D 7 (1973) 2850}.
\bibitem{Davies}
P. Davies, \href{https://iopscience.iop.org/article/10.1088/0305-4470/8/4/022}{J. Phys. A 8 (1975) 609}.
\bibitem{DeWitt}
B. S. DeWitt, General Relativity: An Einstein Centenary Survey, eds S. W. Hawking,
W. Israel (Cambridge Univ Press (1979), Cambridge, UK).
\bibitem{Unruh2}
W. G. Unruh and R. M. Wald, \href{https://link.aps.org/doi/10.1103/PhysRevD.29.1047}{Phys. Rev. D 29 (1984) 1047}.
\bibitem{Muller}
R. M\"{u}ller, \href{https://link.aps.org/doi/10.1103/PhysRevD.56.953}{Phys. Rev. D 56 (1997) 953}.
\bibitem{Vanzella}
D. A. T. Vanzella  and G. E. A.  Matsas, \href{https://link.aps.org/doi/10.1103/PhysRevLett.87.151301}{Phys. Rev. Lett. 87 (2001) 151301}.
\bibitem{Higuichi}
L. C. B. Crispino, A. Higuchi, G. E. A. Matsas, \href{https://link.aps.org/doi/10.1103/RevModPhys.80.787}{Rev. Mod. Phys. 80 (2008) 787}.
\bibitem{Ordonez1}
H. E. Camblong, A. Chakraborty and C. R. Ord\'o\~nez, \href{https://link.aps.org/doi/10.1103/PhysRevD.102.085010}{Phys. Rev. D 102 (2020) 085010}.
\bibitem{Ordonez2}
A. Azizi, H. E. Camblong, A. Chakraborty, C. R. Ord\'o\~nez and M. O. Scully, \href{https://link.aps.org/doi/10.1103/PhysRevD.104.065006}{Phys. Rev. D 104 (2021) 065006}.
\bibitem{Ordonez3}
A. Azizi, H. E. Camblong, A. Chakraborty, C. R. Ord\'o\~nez and M. O. Scully, \href{https://link.aps.org/doi/10.1103/PhysRevD.104.084086}{Phys. Rev. D 104 (2021) 084086}.
\bibitem{Ordonez4}
A. Azizi, H. E. Camblong, A. Chakraborty, C. R. Ord\'o\~nez and M. O. Scully, \href{https://link.aps.org/doi/10.1103/PhysRevD.104.084085}{Phys. Rev. D 104 (2021) 084085}.
\bibitem{ROVELLI}
C. Rovelli, \href{https://doi.org/10.12942/lrr-1998-1}{Living. Rev. Relativ. 1 (1998) 1}.
\bibitem{CARLIP}
S. Carlip, \href{https://iopscience.iop.org/article/10.1088/0034-4885/64/8/301}{Rep. Prog. Phys. 64 (2001) 885}.
\bibitem{ACV}
D. Amati, M. Ciafaloni and G. Veneziano, \href{https://doi.org/10.1016/0370-2693(89)91366-X}{Phys. Lett. B 216 (1989) 41}.
\bibitem{KOPAPR}
K. Konishi, G. Paffuti and P. Provero, \href{https://doi.org/10.1016/0370-2693(90)91927-4}{Phys. Lett. B 234 (1990) 276}.
\bibitem{Reuter0}
M. Reuter, \href{https://link.aps.org/doi/10.1103/PhysRevD.57.971}{Phys. Rev. D 57 (1998) 971}.
\bibitem{Wetterich}
C. Wetterich, \href{https://www.sciencedirect.com/science/article/pii/037026939390726X}{Phys, Lett. B 301 (1993) 90}.
\bibitem{ReuterWetterich}
M. Reuter and C. Wetterich, \href{https://www.sciencedirect.com/science/article/pii/0550321394905436}{Nucl. Phys. B 417 (1994) 181}.
\bibitem{SReuter}
F. Saueressig and M. Reuter, \href{https://doi.org/10.1017/9781316227596}{``\textit{Quantum  Gravity and the Functional Renormalization Group: The Road towards Asymptotic Safety (Cambridge monographs on mathematical physics)} "}, Cambridge University Press (2019).  
\bibitem{Reuter}
A. Bonanno and M. Reuter, \href{https://link.aps.org/doi/10.1103/PhysRevD.62.043008}{Phys. Rev. D 62 (2000) 043008}.

\bibitem{Rohrlich}
F. Rohrlich, \href{https://doi.org/10.1016/0003-4916(63)90051-4}{Ann. Phys. 22 (1963) 169}.

\bibitem{Vallisneri}
M. Pauri and M. Vallisneri, \href{https://doi.org/10.1023/A:1018821619763}{Foun. Phys. 29 (1999) 1499}.

\bibitem{Singleton}
D. Singleton and S. Wilburn, \href{https://link.aps.org/doi/10.1103/PhysRevLett.107.081102}{Phys. Rev. Lett. 107 (2011) 081102}.
\bibitem{Singleton2}
D. Singleton and S. Wilburn, \href{https://doi.org/10.1142/S0218271816440090}{Int. J. Mod. Phys. D 25 (2016) 1644009}.

\bibitem{Fulling}
A. A. Svidzinsky, S. J. Ben-Benjamin, S. A. Fulling and D. N. Page, \href{https://link.aps.org/doi/10.1103/PhysRevLett.121.071301}{Phys. Rev. Lett 121 (2018) 071301}.

\bibitem{Fulling0}
S. A. Fulling and J. H. Wilson, \href{https://iopscience.iop.org/article/10.1088/1402-4896/aaecaa}{Phys. Scr. 94 (2019) 014004}.


\bibitem{SG}
R. Chatterjee, S. Gangopadhyay and A. S. Majumdar, \href{https://link.aps.org/doi/10.1103/PhysRevD.104.124001}{Phys. Rev. D 104 (2021) 124001}.

\bibitem{ParthaKaul}
R. K. Kaul and P. Majumdar, \href{https://link.aps.org/doi/10.1103/PhysRevLett.84.5255}{Phys. Rev. Lett. 84 (2000) 5255}.
\bibitem{Boulware}
D. G. Boulware, \href{https://link.aps.org/doi/10.1103/PhysRevD.11.1404}{Phys. Rev. D 11 (1975) 1404}.
\end{thebibliography}
\end{document}